\documentclass[12pt]{article}
\usepackage{psfig}

\advance\textheight by\topskip

\begin{document}

\title{The problem of identifying decametric sources}

\author{O.V.Verkhodanov$^1$,~H.Andernach$^2$,~N.V.Verkhodanova$^1$
\and {\small $^1$ Special Astrophysical Observatory, Nizhnij Arkhyz,
Russia, 357147}
\and {\small $^2$ Depto.\,de Astronom\'\i a, Apdo.\ Postal 144,
    Univ.\ Guanajuato, Guanajuato, Mexico
}
\and {\it E-mail: vo@sao.ru, heinz@astro.ugto.mx, nat@sao.ru}
}
\date{}
\maketitle

\begin{abstract}
We describe a method to construct continuum spectra for radio sources
on the basis of entries from various source catalogues in comparatively
large error boxes around a given sky position.  Sources from the UTR-2
catalogue (Braude et al.\ 1978--1994), observed at decametric wavelengths
(10--25\,MHz) with an antenna beam of about 40$'$, were
cross-identified with entries from other radio catalogues at higher
frequencies. Using the CATS database
we extracted all sources within 40$'$ around
UTR positions to find candidate identifications.  A spectrum for each source
was fitted with a set of curves using a least-squares method to find
the best fit. We preferentially selected radio counterparts whose
radio spectrum extrapolated to low frequencies matched the UTR decametric 
flux densities, and whose coordinates
were close to the gravity center of UTR positions. Among all the 1822
sources in the UTR catalogue we found about 350 sources to be blends of 
two or more sources. As the most probable true coordinates of the radio
counterparts we used positions from the NVSS, TXS, GB6, or PMN catalogues.
Using low-frequency sources (26, 38, 85\,MHz) from CATS we checked the 
reliability of some of our IDs. We show examples of the above methods, 
including raw and ``cleaned'' spectra.
\end{abstract}

\section{Introduction}

The catalog of 1822 radio sources obtained with the UTR telescope 
near Kharkov (Braude et al.\ 1978--1994, {\tt http://www.ira.kharkov.ua/UTR2/})
covers about 30\% of the sky at six frequencies from 10 to 25~MHz,
and is currently the lowest-frequency catalog of its size.
It provides an ideal basis to study the little known
optical identification content of sources selected at decametric
frequencies. The optical identification rate in the original 
version of the UTR-2 catalog (UTR in what follows) is only 19\%. 
Our goal is to identify all UTR sources with known radio sources.
This cross-identification yields both the radio continuum spectrum
and accurate coordinates for the radio counterparts, thus allowing
to search for optical counterparts on the Digitzed Sky Surveys.
The radio spectral characteristics may also be used to select 
certain samples, e.g. steep radio spectra and the lack of an 
optical identification tend to point to high-redshift radio galaxies.
 
The very large uncertainties of the UTR source positions 
($\sim$0.7$^{\circ}$) forced us to use an interactive process 
to derive radio source spectra.  In this process we use the radio sources
known from low- and intermediate-frequency catalogues as the most likely
candidates, which help us to discard the multitude of weaker sources in
the UTR error box provided e.g. by the recent and sensitive NVSS and
FIRST source catalogues.  All catalog entries are extracted from the 
catalog collection combined in the CATS database (Verkhodanov et al.,
1997).

\section{Input Catalogs}

Characteristics of the main catalogs, used in identification,
are given in Table~2.\\
\begin{center}
\begin{tabular}{|lrcrl|}
\hline
Name   & Freq. &  HPBW(')   & S$_{lim}$(mJy) &  Reference \\
\hline
6C     &    151 &  4.2       & $\sim$200 & 1993MNRAS.263...25Hales+       \\
7C     &    151 &  1.2       &     80    & 1990MNRAS.246..110McGilchrist+ \\
MIYUN  &    232 &  3.8       &$\sim$100  & 1997A\&AS..121...59Zhang+      \\
TXS    &    365 &$\sim$0.1   &$\sim$200  & 1996AJ....111.1945Douglas+     \\
B3     &    408 &3$\times$5  &  100      & 1985A\&AS...59..255Ficarra+    \\
WB92   &   1400 &10$\times$11&  150      & 1992ApJS...79..331White+       \\
87GB   &   4850 &  3.7       &   25      & 1991ApJS...75.1011Gregory+     \\
GB6    &   4850 &  3.7       &   15      & 1996ApJS..103..427Gregory+     \\
PMN    &   4850 &  4.2       &   30      & 1996ApJS..103..145Wright+      \\
MSL    &   misc.& misc.      & misc.     & 1970ApJS...20....1Dixon        \\
\hline
\end{tabular}
\end{center}

\section{Construction of Radio Continuum Spectra}

To prepare radio continuum spectra for decametric sources of the 
UTR catalog (Braude et al., 1978--1994), detected at 10, 12.6, 
14.7, 16.7, 20, and 25~MHz, we first need to identify the sources 
with other known radio sources within their large error boxes 
(we used a box of 40$'\times\,$40$'$) drawn from the CATS database.
CATS provides a graphical interface which displays a ``radio spectrum''
for all sources found in the error box at various frequencies.
By human interaction the most deviant flux measurements in the spectrum
can be recognized as an inappropriate counterpart, and is discarded 
from the spectrum. This ``cleaning'' is achieved 
with the program {\it spg} (Verkhodanov, 1997).

\newpage
The steps used in the identification of UTR sources 
are the following\,:
\begin{enumerate}
\item
    From the main radio catalogs of the CATS database 
   (Verkhodanov et al., 1997), we extract all entries in the search 
   box of 40$'\times\,$40$'$, except for the recent and very sensitive 
  NVSS and FIRST catalogs.
\item
    All objects with flux measurements at several frequencies, are
    separated in the search box of 40$'\times\,$40$'$.
\item
    The spectrum of each object, excluding the UTR data points, is 
   fitted with one of several curves and extrapolated to the UTR frequencies.
\item
    Inside the search box we select counterparts by the 
   following rules: \\[-4.5ex]
    \begin{enumerate}
    \item [(a)]
    the decametric flux densities, as extrapolated from the fitted spectra,
    should be close to the observed UTR fluxes; 
    \item [(b)]
    positions of the radio counterparts should be close to the mean
    position as listed in the UTR catalog.
    \end{enumerate}
    The resulting number of candidate identifications per UTR source 
    ranges from 1 to 4 (see also Fig.~1). In the case of more than one counterpart,
    we consider that all counterparts satisfying the described criteria
    contribute to the UTR source flux, i.e. the UTR detection is a result 
    of blending of one or more independent sources. \\[1ex]

    \begin{figure}[hb!]
    \centerline{
    \hbox{
    \psfig{figure=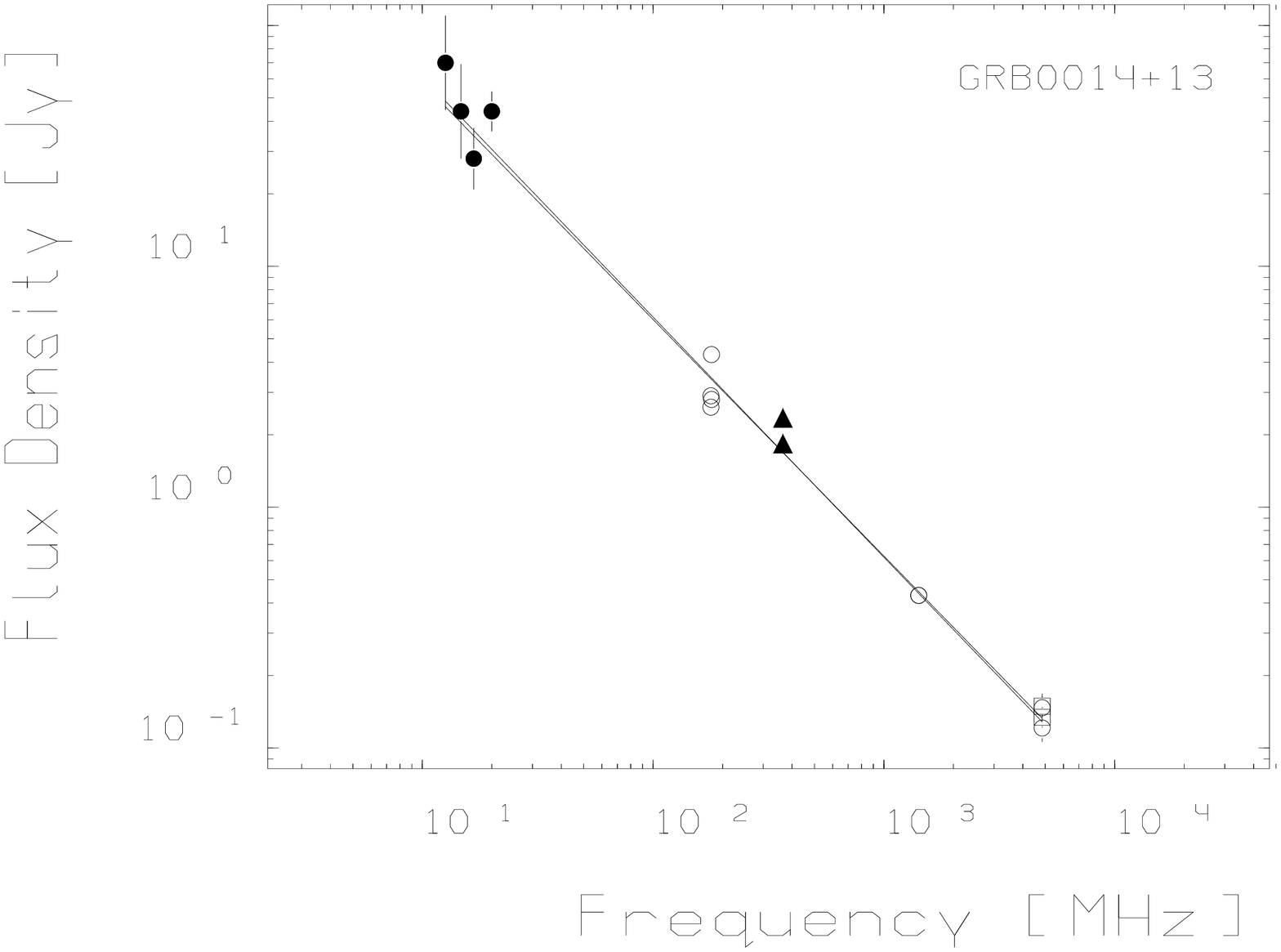,width=8cm,height=6cm,angle=-90}
    \hspace*{0.7cm}
    \psfig{figure=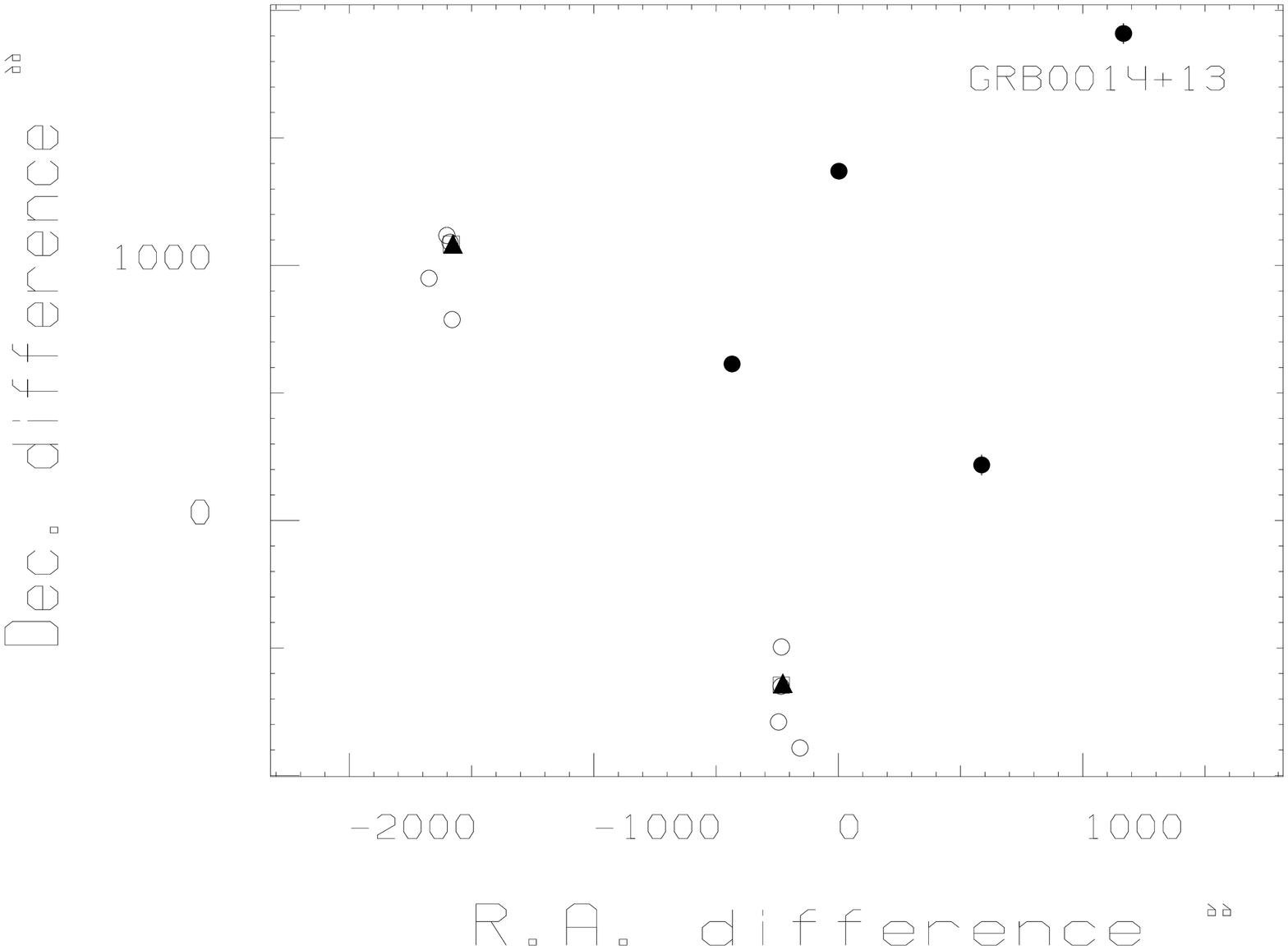,width=7cm,height=6cm,angle=-90}
    }}
    \caption{
     Left panel: spectra of two sources contributing to one UTR object.
     Their contribution in the radio spectrum is practically indistinguishable.
     The right panel shows the relative positions of all entries
     contributing to the spectrum. Two clusters (corresponding to 2 blending sources)
     are visible at the lower and left edge of the chart. The filled dots 
     correspond to UTR fluxes and positions, respectively. 
     The data have already been cleaned from other irrelevant sources in the area.
     The filled sources are UTR points.}
    \end{figure}

\newpage

\item
    For further (e.g.\ optical) identification we defined ``best radio
    coordinates'' from the following catalogs (in order of decreasing 
    priority): TXS (365~MHz), GB6 (4850~MHz), 87GB (4850~MHz), PMN (4850~MHz).
    Data from at least one of these catalogs are included in the 
    information on radio counterparts inside the search box.
\item
    If the identification area is poor of objects (e.g. at low declination,
covered by only few radio surveys), and there are no sources detected 
   simultaneously at several frequencies (i.e.\ no spectral fit was possible), 
   then all objects within the box were retained for further study.
\item
    The best radio coordinates were then used for identification with NVSS or
    FIRST sources. Using flux densities  from NVSS or FIRST usually improved
    the smoothness of the radio spectra.
\item
  Only if an NVSS identification was found, the ``best radio positions'' 
   of item 5 were overwritten with the NVSS position.
\item
    The ``best radio positions'' are used for identification of UTR objects 
   with optical object catalogues (e.g.\ from the APM scans of POSS or ESO/SERC
   surveys) or catalogs in other wavelength ranges.
\end{enumerate}

One of problems of identification inside a wide antenna beam is 
source confusion when one real source has more than one sidelobe.
We have considered such sources and mark them as 'conf' in Table~1.
An example of one such source is shown in Fig.~2. \\

\begin{figure}[h!]
\centerline{
\hbox{
\psfig{figure=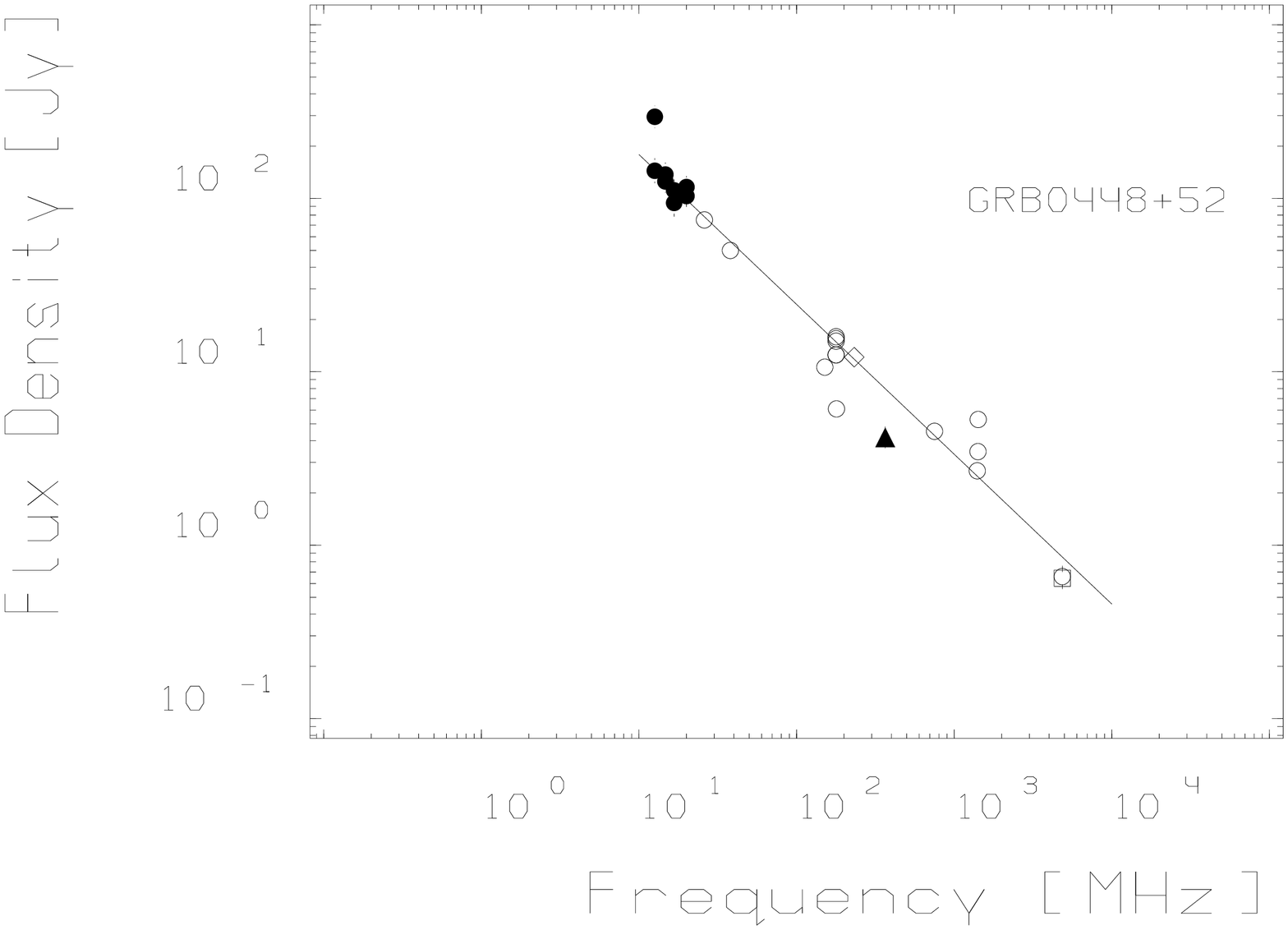,width=8cm,height=6cm,angle=-90}
\hspace*{0.7cm}
\psfig{figure=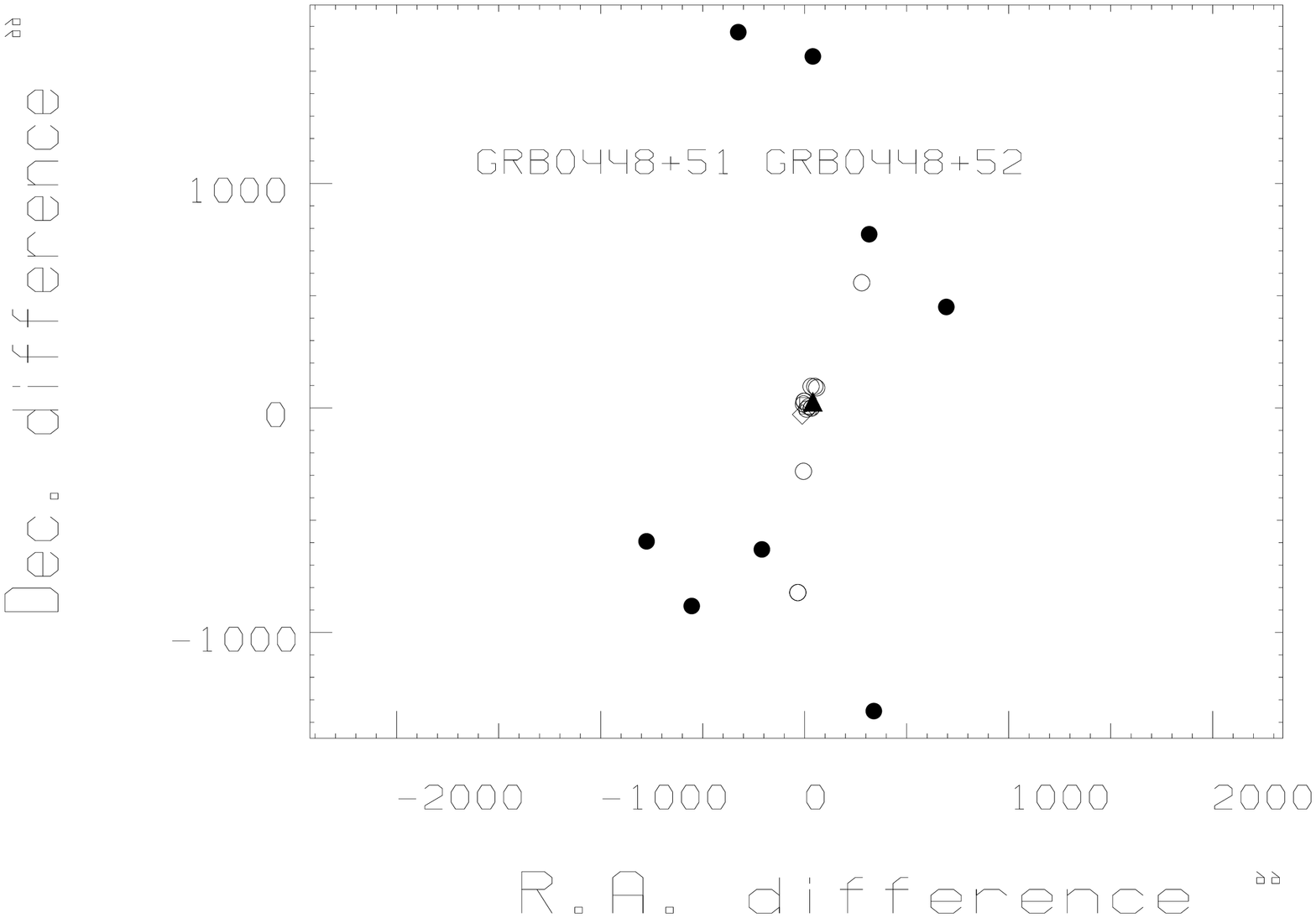,width=7cm,height=6cm,angle=-90}
}}
\caption{
 Example of a real source (4C~52) which has been observed as two 
   independent UTR sources in adjacent observational strips.
 The left panel shows a spectrum, and the right panel shows the
   position of a confused source.
}
\end{figure}

\noindent
To check the reliability of the derived spectra we used the 
low frequency catalogs 6C, 7C (151~MHz), 3C,4C (178~MHz), 
and other catalogs included in Dixon's Master List, like
CL (Viner \& Erickson 1975, 26~MHz), WKB (Williams, Kenderdine \& Baldwin 
1966, 38~MHz), MSH (Mills, Slee \& Hill 1958--61, 85~MHz).
Although these do not cover the entire UTR survey area, they confirm
the high reliability of our methods in the region where these surveys 
overlap.

\section{The Catalog of Counterparts}

The resulting catalog of 2314 radio counterparts, including all blends, 
is given in Table~1.
In the columns of this table we give the original (B1950-based) UTR 
source name, the R.A. and DEC (J2000.0) of the best radio position, 
galactic longitude and latitude, best-fitting radio spectral parameters, 
presence of an optical, infrared or X-ray counterpart, and other names.

Only for three sources we were unable to find identifications 
among the catalogs described above: GR0801$-$11, GR0930$-$00, GR1040$-$02.

There are not enough data to be sure of the identifications of the
sources GR0520$-$08q, GR0537$-$00, GR0629+02, and GR2345+03.

We worked with both existing versions of the UTR catalog:
the printed version  (Braude et al., 1978, 1979, 1981, 1985, 1994)
and the more recent electronic one ({\tt http://www.ira.kharkov.ua/UTR2/}).
We included UTR sources of all reliability levels (A,B,C) as given in
the UTR catalog.

In Table~1 we kept the names from the printed catalog version for the 
following sources (new names from the electronic version are given in brackets): \\
{\small
GR0224+03 ~(GR0227+03),~~
GR0307+17 ~(GR0307+16),~~
GR0411+14 ~(GR0411+13),~~ \\
GR0919+55 ~(GR0918+55),~~
GR0929+07 ~(GR0930+07),~~
GR1039+03 ~(GR1039+02),~~ \\
GR1142+00 ~(GR1142$-$00),~~
GR1538+01 ~(GR1539+01),~~
GR1547+03 ~(GR1548+03).~~
}

\medskip
Table~1 also includes the following sources, which are present in the printed version,
but are absent in electronic one\,: \\
  GR1915+56,
  GR2355+19,
  GR2355$-$02,
  GR2358+08.

\section {Conclusion}

This method including simultaneous account of spectra behavior and
radio points concentration on the coordinates plain is a good solution for
the problem of the identification radio sources observed with a large
antenna beam.

Different subsamples of sources identified with the above methods are
being studied by the authors.

The authors are grateful to our collegues Miroshnichenko A.P. and Krivitskij D.
from the Institute of Radio Astronomy (Kharkov) for providing data and
useful discussion. We also thank Sergej Trushkin and Vladimir Chernenkov 
(SAO RAS), co-creators of the CATS database, for fruitful discussions.

\end{document}